\documentclass[sigconf]{acmart}
\AtBeginDocument{%
  \providecommand\BibTeX{{%
    Bib\TeX}}}

\setcopyright{acmlicensed}
\copyrightyear{2025}
\acmYear{2025}
\setcopyright{rightsretained}
\acmConference[RecSys GenAIECommerce'25]{Proceedings of the 18th International ACM Conference on Recommender Systems}{September 22, 2025}{Prague, Czech Republic}
\acmBooktitle{Proceedings of the 2025 RecSys Workshop on Generative AI for E-commerce (RecSys GenAIECommerce'25), September 22, 2025, Prague, Czech Republic}
\acmDOI{}

\acmSubmissionID{209}

\usepackage[skip=0pt]{caption}
\usepackage{booktabs}
\usepackage{numprint}
\npthousandsep{,}
\npdecimalsign{.}
\usepackage[inline]{enumitem}
\usepackage{balance}

\usepackage{bm}
\usepackage{bbm}

\usepackage{xcolor}

\usepackage{verbatim}

 \usepackage{multirow}

\usepackage{graphicx} %
\graphicspath{{img/}} %

\usepackage{acronym}
\acrodef{IR}{information retrieval}

\usepackage{subcaption}

\usepackage{tikz}
\usetikzlibrary{shapes.geometric, arrows.meta, positioning}

\tikzstyle{process} = [rectangle, rounded corners, minimum width=3.2cm, minimum height=1.2cm, text centered, draw=black, fill=blue!10]
\tikzstyle{input} = [trapezium, trapezium left angle=70, trapezium right angle=110, minimum width=2.8cm, minimum height=1cm, text centered, draw=black, fill=green!10]
\tikzstyle{arrow} = [thick, ->, >=stealth]

\newcommand{\header}[1]{\vspace{1mm}\noindent\textbf{#1}.}

\AtBeginDocument{%
  \providecommand\BibTeX{{%
    \normalfont B\kern-0.5em{\scshape i\kern-0.25em b}\kern-0.8em\TeX}}}

\newcommand{\ourtask}{Image-Seeking Intent Prediction}

\newcommand{\OurModel}{IRP}

\acrodef{SOTA}{state-of-the-art}
\acrodef{IR}{information retrieval}
\acrodef{FP}{false positive}
\acrodef{VUI}{voice-user interface}
\acrodef{VA}{virtual assistant}
\acrodef{ISIP}{image-seeking intent prediction}
\acrodef{MMR}{Maximal Marginal Relevance}

\begin{document}

\title{Image-Seeking Intent Prediction for Cross-Device Product Search}

\author{Mariya Hendriksen}
\affiliation{%
  \institution{University of Oxford}
  \country{United Kingdom}
}
\authornote{Work done while interning at Amazon.}

\author{Svitlana Vakulenko}
\affiliation{%
  \institution{Vienna University of Economics and Business}
  \country{Austria}
}

\author{Jordan Massiah}
\affiliation{%
  \institution{Amazon}
  \country{United Kingdom}
}

\author{Gabriella Kazai}
\affiliation{%
  \institution{Amazon}
  \country{United Kingdom}
}

\author{Emine Yilmaz}
\affiliation{%
  \institution{Amazon}
  \country{United Kingdom}
}

\renewcommand{\shortauthors}{Hendriksen et al.}
\acmArticleType{Research}
\begin{abstract}
Large Language Models (LLMs) are transforming personalized search, recommendations, and customer interaction in e-commerce. Customers increasingly shop across multiple devices, from voice-only assistants to multimodal displays, each offering different input and output capabilities. A proactive suggestion to switch devices can greatly improve the user experience, but it must be offered with high precision to avoid unnecessary friction.
We address the challenge of predicting when a query requires visual augmentation and a cross-device switch to improve product discovery. We introduce Image-Seeking Intent Prediction, a novel task for LLM-driven e-commerce assistants that anticipates when a spoken product query should proactively trigger a visual on a screen-enabled device. Using large-scale production data from a multi-device retail assistant, including 900K voice queries, associated product retrievals, and behavioral signals such as image carousel engagement, we train IRP (Image Request Predictor), a model that leverages user input query and corresponding retrieved product metadata to anticipate visual intent.
Our experiments show that combining query semantics with product data, particularly when improved through lightweight summarization, consistently improves prediction accuracy. Incorporating a differentiable precision-oriented loss further reduces false positives. These results highlight the potential of LLMs to power intelligent, cross-device shopping assistants that anticipate and adapt to user needs, enabling more seamless and personalized e-commerce experiences.
\end{abstract}

\keywords{image-seeking intent, virtual assistants, generative models, cross-device search}

\maketitle

\section{Introduction}
\label{sec:introduction}

Large Language Models (LLMs) have achieved state-of-the-art performance across a variety of tasks, including text generation, reasoning, and question answering \citep{brown2020language, chowdhery2022palm, openai2023gpt4}. Their integration into virtual assistants (VAs) offers new opportunities for intelligent, context-aware interaction \citep{gondala2021error, sannigrahi2024synthetic, van2020predicting}. In the e-commerce domain, these assistants operate within a diverse ecosystem of devices, ranging from voice-only smart speakers to screen-enabled smartphones and tablets \citep{zhang2024server, bai2023natural, DBLP:journals/sigir/TsagkiasKKMR20}. Each device supports distinct input and output modalities, and product catalogs increasingly contain multimodal representations that include visual, textual, and structured data. These assistants must retrieve and present such information in modality-appropriate ways. As customer expectations for personalized and visually rich recommendations continue to grow, the ability of LLM-driven assistants to adapt content presentation to device capabilities is becoming a critical requirement.

A growing challenge in this setting is \textit{proactive device switching}, which involves determining when a query issued to a screenless assistant requires a visual response that would be better served on a screen-enabled device. Prior research in cross-device search has primarily focused on predicting the next device a user will access. In contrast, our objective is to determine when a device switch should be suggested to better serve the user’s intent. This distinction is particularly important when the user’s current device lacks the modalities necessary to meet the request, such as the ability to display product images or other forms of visually enhanced recommendations.

We address this challenge through Image-Seeking Intent Prediction, a novel task for LLM-driven e-commerce assistants. The goal is to predict whether a user query, issued via speech on a screenless device, will require a visual modality to fulfill the user’s need. This problem arises when a voice-only assistant must decide if presenting images or other visual content would improve the experience. Once visual intent is detected, the assistant can proactively suggest switching to a screen-enabled device or adapt the interface accordingly.
We cast this task as a binary, highly imbalanced classification problem, where a positive label indicates that the current voice-only interaction should be escalated to a screen-based device. The task is motivated by two key use cases: (i) enhancing multi-device user experiences by recommending a switch to a more appropriate modality \citep{montanez2014cross, han2015understanding}; and (ii) supporting adaptive user interfaces that respond intelligently to predicted modality needs, for example, by automatically showing images when necessary \citep{oliveira2023evolution}. The problem presents several challenges: (i) Minimizing false positives — unnecessary device switches disrupt the user experience, making high precision essential; (ii) Label imbalance — device-switch events are rare, and thus switch-worthy queries form only a small fraction of the data; and (iii) Multimodal context — utterances are spoken, while retrieved products contain a mixture of textual, structured, and visual content.

\header{Dataset}  
Direct supervision for device-switching behavior is rarely available, and manually annotating visual intent at scale is prohibitively expensive. We address this by defining a proxy task that leverages behavioral signals on screen-enabled devices. Specifically, we use the action of tapping to open the image carousel as a weak but scalable indicator of visual intent. This interaction enables large-scale supervision from real-world data without manual labeling. The model is trained to predict, given an utterance and retrieved products, whether the user would have activated the image carousel, signaling a transition to a visual modality.
We construct a large-scale dataset of \numprint{2.2}M user interactions collected over six months from a production-grade multi-device assistant system. The dataset includes over \numprint{900000} spoken utterances paired with the top-10 retrieved products per query, along with behavioral labels from downstream interaction signals.

\header{Image Request Predictor}  
To address \ourtask{}, we propose \OurModel{}, a model with a generative backbone that predicts image-seeking intent from the user’s utterance and associated retrieved product information. \OurModel{} jointly encodes the utterance and product summaries to make predictions about visual needs.
We compare \OurModel{} against strong baselines using only utterance transcriptions, intent classifications, or raw product titles. We find that combining utterance data with structured, summarized product information yields significant performance gains.

\noindent
We answer the following research questions:
\begin{enumerate}[label=(RQ\arabic*)]
	\item How does the integration of utterance-level and product-level features influence predictive performance in \ourtask{}?
	\item What is the effect of representing product information through lightweight summarization compared to using raw product titles or attributes?
	\item To what extent does optimizing the loss function explicitly for precision influence performance in this high-precision, imbalanced classification setting?
\end{enumerate}

The principal contributions of our research are the following:
\begin{enumerate}[label=(\roman*)]
    \item We define Image-Seeking Intent Prediction, a novel task for language model-based cross-device assistants in e-commerce, and propose the Image Request Predictor (IRP), a model for the task.
    \item We conduct extensive evaluation on a large-scale proprietary dataset of real user interactions in a multi-device e-commerce assistant system.
    \item We discover that combining query semantics with summarized product data yields consistent accuracy gains, while a differentiable precision-oriented loss further reduces false positives.
\end{enumerate}

\section{Methodology}
\label{sec:approach}

We formalize the \ourtask{} task and present \OurModel{}, an architecture with three modules: an utterance processing pipeline, a product information processing pipeline, and an image request prediction module. Together, these components learn to determine whether a spoken query requires switching from a voice-only to a screen-enabled device.

\subsection{Task definition}
Following prior work \citep{varamesh2020self, brown2020smooth, zhang2022contrastive}, we represent each user interaction as a pair $(U, \mathcal{P})$, where $U$ is the user query and $\mathcal{P} = {\mathbf{p}^{(1)}, \dots, \mathbf{p}^{(k)}}$ is the set of the top-$k$ products retrieved in response to $U$. Each query $U$ consists of a transcription $u_{\text{q}}$ and an associated intent label $u_{\text{int}}$. Each retrieved product $\mathbf{p}^{(j)} \in \mathcal{P}$ is described by a structured set of attributes:
\begin{align}
\mathbf{p}^{(j)} = \{ & p^{(j)}_{\text{title}},\ p^{(j)}_{\text{brand}},\ p^{(j)}_{\text{size}},\ p^{(j)}_{\text{color}},\ p^{(j)}_{\text{reviews}}, \notag \\
                      & p^{(j)}_{\text{price}},\ p^{(j)}_{\text{style}},\ p^{(j)}_{\text{group}},\ p^{(j)}_{\text{type}} \}
\end{align}

Due to the lack of labeled supervision for device switching events and the high cost of manual annotation, we adopt a scalable proxy task. Specifically, we use the user’s interaction with an image carousel (i.e., whether they tap to open it) as an implicit signal of visual intent. This assumption enables the training of models at scale using real behavioral data.

\begin{figure}[t]
\centering
\includegraphics[clip,trim=0mm -5mm 0mm 0mm,width=0.85\linewidth]{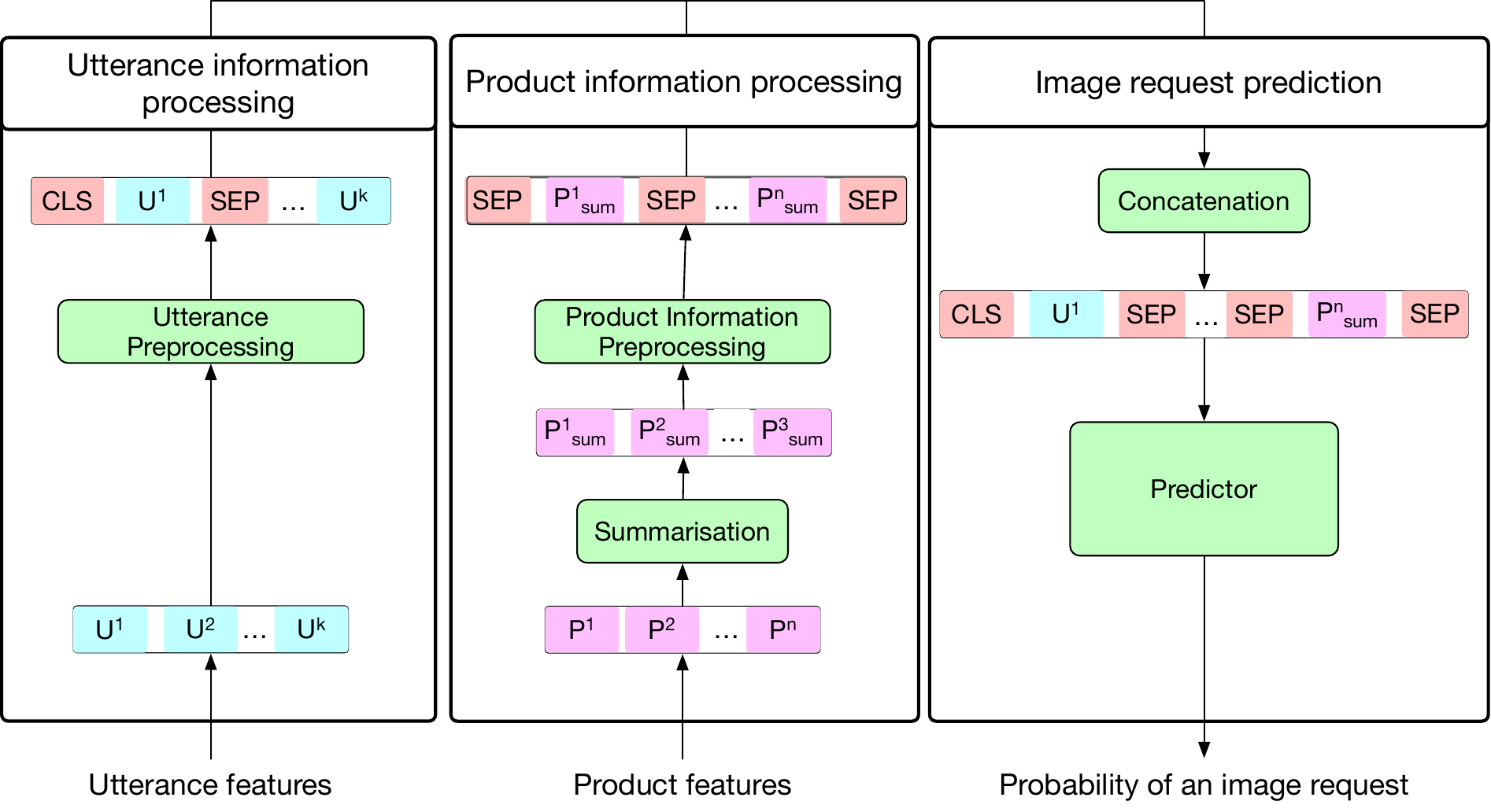}
\caption{\OurModel{} architecture. The query and product information are processed through respective pipelines, fused into a joint representation, and passed to a prediction head that estimates the likelihood of the image-seeking intent.
}
\label{fig:our-approach}
\end{figure}

\subsection{Image Request Predictor}

Figure~\ref{fig:our-approach} illustrates the architecture of \OurModel{}. The model processes utterance and product information through separate pipelines, summarizes retrieved product representations, and fuses them to predict whether the user’s request implies a need for visual content.

\paragraph{Utterance Processing Pipeline}  
The user query $U$ is transcribed into text and passed to a preprocessing function $f_{\text{up}}$, which converts it into a tokenized sequence:
\begin{equation}
\text{seq}_u = f_{\text{up}}(U) = ([\text{CLS}], u^1, [\text{SEP}], \dots, u^K, [\text{SEP}]),
\end{equation}
where $[\text{CLS}]$ and $[\text{SEP}]$ are standard special tokens.

\paragraph{Product Information Processing Pipeline}  
The retrieved products $\mathcal{P}$ are summarized using a function $f_{\text{sum}}$, which produces an aggregated vector:
\begin{equation}
\mathbf{p}_{\text{sum}} = f_{\text{sum}}(\mathcal{P}) = \frac{1}{k} \sum_{j=1}^{k} \mathbf{p}^{(j)}.
\end{equation}
This summary is then tokenized via a product preprocessing function $f_{\text{pp}}$:
\begin{equation}
\text{seq}_p = f_{\text{pp}}(\mathbf{p}_{\text{sum}}) = ([\text{SEP}], p^1, [\text{SEP}], \dots, p^N, [\text{SEP}]).
\end{equation}

\paragraph{Fusion and Prediction}  
The utterance and product sequences are integrated by a fusion module $f_{\text{fusion}}$, parametrized to operate either as direct sequence concatenation or as concatenation over a reduced set of product tokens obtained via Maximal Marginal Relevance (MMR) selection \cite{carbonell1998use}. Formally:
\begin{align}
X &= f_{\text{fusion}}(\text{seq}_u, \text{seq}_p) \\
  &= ([\text{CLS}], u^1, \dots, u^K, [\text{SEP}], p^1, \dots, p^N, [\text{SEP}]),
\end{align}
where $(p^1, \dots, p^N)$ denotes either all product tokens or the MMR-selected subset.  
The fused representation $X$ is then passed through a classification head $f_{\text{pred}}$ to produce the probability of image-seeking intent:
\begin{equation}
p = f_{\text{pred}}(X), \quad 0 \leq p \leq 1.
\end{equation}

\subsection{Training Objectives}

We model \ourtask{} as binary classification and evaluate three loss formulations.

\paragraph{Binary Cross-Entropy (BCE)}  
The BCE loss minimizes cross-entropy between predicted probability $p$ and label $y \in \{0,1\}$:
\begin{equation}
\mathcal{L}_{\text{BCE}} = -y \log(p) - (1 - y) \log(1 - p).
\end{equation}

\paragraph{Precision Loss}  
To penalize false positives explicitly, we use a differentiable approximation of precision~\citep{benedict2021sigmoidf1, patel2022recall}:
\begin{equation}
\mathcal{L}_{\text{Precision}} = 1 - \frac{\text{TP}}{\text{TP} + \text{FP}},
\end{equation}
where $\text{TP} = \sum \hat{y} \cdot y$ and $\text{FP} = \sum \hat{y} \cdot (1 - y)$, with $\hat{y}$ as the model’s sigmoid output.

\paragraph{Combined Loss}  
To balance general accuracy with false-positive reduction, we define:
\begin{equation}
\mathcal{L}_{\text{Sum}} = \alpha \, \mathcal{L}_{\text{BCE}} + \beta \, \mathcal{L}_{\text{Precision}},
\end{equation}
where $\alpha$ and $\beta$ are hyperparameters controlling the trade-off between optimizing overall classification accuracy and prioritizing high precision.

\section{Experiments}
\label{sec:experiments}
We empirically evaluate the \ourtask{} formulation and \OurModel{} on large-scale real-world interaction data from a production conversational e-commerce assistant. We describe the dataset, model backbones, and evaluation metrics, then report results for feature contributions, product summarization, precision-oriented loss functions, and backbone comparisons, followed by implementation details.

\subsection{Dataset}
We construct a large-scale dataset from de-identified interaction logs collected between August and December 2022 from a production conversational e-commerce assistant. The system supports both screenless devices (e.g., smart speakers) and screen-enabled devices (e.g., mobile phones, tablets), making it a natural setting for studying cross-modality behavior.

\paragraph{Data Sampling and Filtering}
We start with \numprint{2200000} user interactions sampled over five months, containing both utterance data and downstream behavioral signals such as taps. Approximately 10\% of these interactions include an image gallery opening action. To focus on voice-initiated interactions, we filter for utterances explicitly triggered by the user. This yields a high-quality subset of \numprint{900000} utterance-based interactions, in which 20\% are followed by an image gallery opening. This proportion provides a sufficiently large and meaningful set of positive examples for our proxy task.

\paragraph{Dataset Characteristics}
The resulting dataset reflects realistic user behavior in a high-traffic, multi-device e-commerce assistant. Each instance in $\mathcal{D}$ is represented as a tuple $\left(U^i, \mathcal{P}^i, y^i\right)$, where $U^i$ is a spoken utterance with both transcription and intent; $\mathcal{P}^i = {\mathbf{p}^{(i,1)}, \dots, \mathbf{p}^{(i,k)}}$ is the set of top-$k$ retrieved products for $U^i$; and $y^i \in {0,1}$ is a binary label indicating whether the user opened the image carousel, serving as a proxy for visual intent. Each product $\mathbf{p}^{(i,j)}$ contains structured metadata including title, brand, size, color, price, reviews, style, and product category. The dataset thus captures:
\begin{enumerate*}[label=(\roman*)]
\item natural language voice queries across diverse product search contexts,
\item real-world retrieval results from a production search system, and
\item implicit behavioral signals indicative of multimodal information needs.
\end{enumerate*}

This configuration enables training and evaluation for the \ourtask{} task at scale, using weak supervision derived from real-world behavioral data in place of costly manual annotation.

\subsection{Model Backbones}  
In all experiments, the Image Request Predictor (IRP) uses transformer-based language model backbones. We evaluate three pretrained architectures:
\begin{itemize}
    \item \textbf{DistilBERT}~\citep{sanh2019distilbert}: A compressed variant of BERT that reduces parameters by ~40\% and achieves up to 60\% faster inference while retaining over 95\% of BERT’s performance. It is trained via knowledge distillation with a composite loss combining language modeling, distillation, and cosine embedding objectives.
    \item \textbf{RoBERTa}~\citep{liu2019roberta}: An optimized variant of BERT that removes Next Sentence Prediction, uses larger mini-batches and learning rates, trains on substantially more data, and applies dynamic masking, resulting in consistent performance improvements over BERT.
    \item \textbf{XLNet}~\citep{yang2019xlnet}: A generalized autoregressive model that captures bidirectional context without input corruption by maximizing expected log-likelihood over all factorization permutations. It incorporates Transformer-XL features such as segment recurrence and relative positional encoding to better handle long-range dependencies.
\end{itemize}

These backbones are integrated into \OurModel{} for encoding the concatenated utterance and product sequences (see Section~\ref{sec:approach}), and we use them consistently throughout training and evaluation. Backbone comparisons are reported in an ablation study later in this section.

\subsection{Metrics} In \ourtask{}, the high cost of unnecessary cross-device switches requires prioritizing \textit{precision}, the proportion of predicted positives that are correct:
\begin{equation}
    P = \frac{\text{TP}}{\text{TP} + \text{FP}} ,
\end{equation}
where TP and FP denote true and false positives, respectively.  
We also report \textit{recall}, the proportion of actual positives correctly identified:
\begin{equation}
    R = \frac{\text{TP}}{\text{TP} + \text{FN}} ,
\end{equation}
where FN denotes false negatives.
To jointly assess both metrics, we use the $F_{\beta}$ score:
\begin{equation}
F_{\beta} = \frac{(1 + \beta^2) \cdot P \cdot R}{\beta^2 \cdot P + R},
\end{equation}
where $\beta$ controls the relative weight of recall. We set $\beta = 0.5$, placing greater emphasis on precision while retaining sensitivity to recall.

\subsection{Experimental Setup and Results}
\label{sec:experimental-setup}
We conduct three experiments to answer the research questions outlined in Section~\ref{sec:introduction}. Each experiment investigates a specific aspect of the model architecture or training strategy. For each, we present the setup, results, and key insights.

\begin{table}[t]
\caption{Feature Contribution Analysis.}
\label{tab:exp1}
\begin{tabular}{lccc}
\toprule
 & \textbf{Precision} & \textbf{Recall} & $\bm{F_{05}}$ \\
\midrule
Utterance & 77.83 & 64.92 & 73.41 \\
Intent & 76.12 & 40.53 & 56.32 \\
Title & 76.59 & 56.76 & 68.90 \\
\midrule
Utterance, intent & 77.93 & 65.11 & 73.55 \\
Intent, title & 77.34 & 57.81 & 69.77 \\
Utterance, title & 77.91 & 63.75 & 72.94 \\
Utterance, intent, title & 78.11 & 64.19 & 73.23 \\
\bottomrule
\end{tabular}
\end{table}

\subsubsection{Feature Contribution Analysis} 
We begin by evaluating the contribution of different feature types to model performance. We test models trained with:
\begin{enumerate*}[label=(\roman*)]
    \item utterance transcription,
    \item utterance intent,
    \item product title (top-1 product),
    \item combinations of the above.
\end{enumerate*}
Utterance features $U$ are tokenized via $f_{\text{up}}$ to produce $\text{seq}_u$, and product titles $P$ are processed via $f_{\text{pp}}$ to produce $\text{seq}_p$. Combined features are fused via $f_{\text{concat}}$ and passed to the image request predictor.

\header{Results}  
Table~\ref{tab:exp1} shows that utterance features alone achieve the best performance across all metrics. However, combining utterance with intent and product title yields even better Precision (78.11) and a strong overall performance. Notably, utterance + intent gives the highest Recall (65.11) and $F_{0.5}$ (73.55).

\header{Upshot}  
Utterance features are the strongest individual signal, but combining them with product metadata enhances performance. Feature fusion significantly improves Precision, which is critical for avoiding unnecessary modality switches.

\begin{table}[t]
\caption{Retrieved Product Summarization.}
\label{tab:exp2}
\begin{tabular}{lccc}
\toprule
 & \textbf{Precision} & \textbf{Recall} & $\bm{F_{05}}$ \\
\midrule
Agg. by mean $(f_{sum})$ & 78.47 & 60.42 & 71.61 \\
MMR~\cite{carbonell1998use} & 78.27 & 61.94 & 72.27\\
\bottomrule
\end{tabular}
\end{table}

\subsubsection{Product Summarization and Representation}
We expand the product representation by including all selected product attributes (title, brand, size, color, reviews, price, style, group, and type). Due to input length constraints, we summarize the top-$k$ retrieved products using:
\begin{enumerate*}[label=(\roman*)]
    \item Mean pooling ($f_{\text{sum}}$),
    \item Maximal Marginal Relevance (MMR)~\cite{carbonell1998use}.
\end{enumerate*}
The summarized vector $\mathbf{P}_{\text{sum}}$ is then tokenized and concatenated with utterance features as in Experiment 1.

\header{Results}  
As shown in Table~\ref{tab:exp2}, aggregation by mean improves Precision (78.47), while MMR yields the highest Recall (61.94) and $F_{0.5}$ (72.27). Both summarization techniques lead to better results than using product titles alone.

\header{Upshot}  
Incorporating a richer product representation improves model performance, especially when combined with effective summarization. Mean pooling favors precision, while MMR offers a better balance between precision and recall.

\subsection{Ablation Studies}
We conduct a series of ablation experiments to isolate the effects of main design choices in \OurModel{}. 

\subsubsection{Ablation on Loss Functions}
We compare three training objectives: (i) $\mathcal{L}_{\text{BCE}}$, the standard binary cross-entropy loss; (ii) $\mathcal{L}_{\text{Precision}}$, a differentiable surrogate for precision; and (iii) $\mathcal{L}_{\text{Sum}}$, a weighted combination of the two. Each loss function is evaluated across multiple input configurations, including single-feature models (e.g., utterance, intent, or title), feature pairs, and the full input setting using all features.

\begin{table}[t]
\caption{Ablation on Loss Functions.}
\label{tab:exp3}
\resizebox{\linewidth}{!}{
\begin{tabular}{llccr}
\toprule
 & \textbf{Loss} & \textbf{Precision} & \textbf{Recall} & $\bm{F_{05}}$ \\
\midrule
 & $\mathcal{L}_{\text{BCE}}$ & 77.83 & 64.92 & 73.41 \\
 & $\mathcal{L}_{\text{Precision}}$ & 77.59 & 65.54 & 73.55 \\
\multirow{-3}{*}{Utterance} & $\mathcal{L}_{\text{Sum}}$ & 77.89 & 64.02 & 73.05 \\
\midrule
 & $\mathcal{L}_{\text{BCE}}$ & 76.12 & 40.53 & 56.32 \\
 & $\mathcal{L}_{\text{Precision}}$ & 76.32 & 31.03 & 42.43 \\
\multirow{-3}{*}{Intent} & $\mathcal{L}_{\text{Sum}}$ & 76.11 & 40.53 & 56.33 \\
\midrule
 & $\mathcal{L}_{\text{BCE}}$ & 76.59 & 56.76 & 68.90 \\
 & $\mathcal{L}_{\text{Precision}}$ & 76.36 & 53.39 & 66.87 \\
\multirow{-3}{*}{Title} & $\mathcal{L}_{\text{Sum}}$ & 76.56 & 56.63 & 68.82 \\
\midrule
 & $\mathcal{L}_{\text{BCE}}$ & 77.93 & 65.11 & 73.55 \\
 & $\mathcal{L}_{\text{Precision}}$ & 77.48 & 65.63 & 73.53 \\
\multirow{-3}{*}{Utterance, intent} & $\mathcal{L}_{\text{Sum}}$ & 77.91 & 63.67 & 72.90 \\
\midrule
 & $\mathcal{L}_{\text{BCE}}$ & 77.34 & 57.80 & 69.77 \\
 & $\mathcal{L}_{\text{Precision}}$ & 77.23 & 52.38 & 66.52 \\
\multirow{-3}{*}{Intent, title} & $\mathcal{L}_{\text{Sum}}$ & 77.09 & 58.52 & 70.04 \\
\midrule
 & $\mathcal{L}_{\text{BCE}}$ & 77.91 & 63.75 & 72.93 \\
 & $\mathcal{L}_{\text{Precision}}$ & 77.64 & 63.64 & 72.75 \\
\multirow{-3}{*}{Utterance, title} & $\mathcal{L}_{\text{Sum}}$ & 77.99 & 64.88 & 73.48 \\
\midrule
 & $\mathcal{L}_{\text{BCE}}$ & 78.11 & 64.19 & 73.23 \\
 & $\mathcal{L}_{\text{Precision}}$ & 77.86 & 62.61 & 72.39 \\
\multirow{-3}{*}{Utterance, intent, title} & $\mathcal{L}_{\text{Sum}}$ & 78.20 & 63.60 & 73.01 \\
\midrule
 & $\mathcal{L}_{\text{BCE}}$ & 78.47 & 60.42 & 71.61 \\
 & $\mathcal{L}_{\text{Precision}}$ & 77.99 & 62.87 & 72.58 \\
\multirow{-3}{*}{\OurModel{}} & $\mathcal{L}_{\text{Sum}}$ & 78.22 & 63.27 & 72.87\\
\bottomrule
\end{tabular}
}
\end{table}

\header{Results}  
Table~\ref{tab:exp3} summarizes the outcomes. Models trained with $\mathcal{L}_{\text{BCE}}$ and $\mathcal{L}_{\text{Sum}}$ achieve the best performance in eleven configurations each, while $\mathcal{L}_{\text{Precision}}$ leads in only three. In most settings, $\mathcal{L}_{\text{Sum}}$ is competitive or superior to the others in $F_{0.5}$.

\header{Upshot}  
While BCE remains a strong baseline, incorporating precision into the loss yields consistent improvements in high-precision use cases. The combined loss strikes a favorable balance, especially for rare-event detection where false positives are costly.

\subsubsection{Ablation on Predictor Backbone}  
To assess the trade-off between performance and model efficiency, we conduct an ablation study evaluating different transformer backbones for the image request predictor module in \OurModel{}. Specifically, we compare three widely used sequence classification architectures: distilBERT~\cite{sanh2019distilbert}, RoBERTa~\cite{liu2019roberta}, and XLNet~\cite{yang2019xlnet}.

\begin{table}[t]
\caption{Ablation on Predictor Backbone.}
\label{tab:ablation-predictor}
\begin{tabular}{lcccr}
\toprule
\textbf{Architecture} & \textbf{Precision} & \textbf{Recall} & $\bm{F_{0.5}}$ & \textbf{\#Params} \\
\midrule
DistilBERT~\cite{sanh2019distilbert} & 78.47 & 60.42 & 71.61 & 66M \\
RoBERTa~\cite{liu2019roberta} & 78.78 & 59.28 & 71.14 & 125M \\
XLNet~\cite{yang2019xlnet} & 78.79 & 60.64 & 71.85 & 340M \\
\bottomrule
\end{tabular}
\end{table}

As shown in Table~\ref{tab:ablation-predictor}, XLNet achieves the best overall performance across all metrics. However, distilBERT performs competitively while using significantly fewer parameters -- just one-fifth the size of XLNet -- highlighting its suitability for deployment in resource-constrained environments.

\subsection{Implementation Details}\label{sec:implementation-details} 
All models are trained for 20 epochs using a batch size of $\beta = 128$, a learning rate of $\eta = 2e\!-\!5$ and a weight decay of $\lambda = 1e\!-\!2$. Optimization is performed using the AdamW optimizer~\cite{loshchilov2017decoupled}. We initialize all models using publicly available pretrained checkpoints~\citep{wolf2019huggingface}.

\section{Related Work}
\label{sec:related-work}

\header{Multimodal Retrieval in E-Commerce}  
mariyaE-commerce retrieval systems increasingly leverage multimodal inputs such as text, images, and structured product metadata for ranking and recommendation~\citep{DBLP:journals/sigir/TsagkiasKKMR20, zhang2022contrastive, hendriksen2022extending, hendriksen2022multimodal, mariya-hendriksen-phd-thesis-2024}. Vision–language models have been used to improve product understanding and retrieval quality~\citep{hosseini2024retrieve, yang2024multimodal, hendriksen2023scene}, while virtual assistants are evolving toward richer multimodal interactions~\citep{oliveira2023evolution}. However, most prior work assumes a fixed modality, focusing on improving retrieval accuracy or dialogue generation.  
In contrast, we address the problem of deciding when a query is best served via a visual modality. Rather than fusing modalities for ranking, we use retrieved product metadata to infer visual content needs, enabling proactive e-commerce assistant behavior.

\header{User Intent Prediction}  
Inferring user intent from implicit behavioral signals is a long-standing topic in information retrieval and recommendation~\citep{joachims2017accurately, agichtein2006improving}. Clicks, hovers, and scrolls have been widely used in learning-to-rank and personalized search. More recent work has modeled multiple latent intents from multimodal data and implicit feedback~\citep{yang2024multimodal}, or inferred visual intent from user actions in conversational settings~\citep{zang2021photochat}.  
Our approach shares the use of implicit feedback but differs in focus and supervision: (i) we predict modality preference rather than item relevance or task intent, and (ii) we employ a specific behavioral signal (image carousel taps) as a scalable, weakly supervised proxy for visual information need, enabling high-precision intent models without manual labeling.

\header{Cross-Device Search}  
Cross-device behavior has been extensively studied in web and mobile search~\citep{montanez2014cross, han2015understanding, wang2013characterizing}, often modeling device transitions within multi-session tasks. Recent work explores proactive adaptation in assistants, where systems adjust to situational context such as attention, environment, or device constraints~\citep{hendriksen-2020-analyzing, deng2024towards, samarinas2024procis}.  
We differ in scope and formulation: instead of predicting the next device, we predict whether a voice-issued query on a screenless device requires a visual response. This prediction can proactively trigger a device-switch suggestion. By combining utterance understanding, product metadata, and weak supervision, our method enables assistants to anticipate and act on latent modality needs in real time.

\section{Conclusions}
\label{sec:conclusions}
We introduced \ourtask{}, the task of predicting when a query to a screenless assistant should trigger a proactive switch to a screen-enabled device. This setting is important for cross-device e-commerce assistants, where some interactions require visual augmentation. Given the high cost of unnecessary switches, we emphasized precision and optimized for both Precision and $F_{0.5}$. We proposed \OurModel{}, which integrates features from user utterances with product data, and improved training with a combined binary cross-entropy and differentiable precision loss, reducing false positives while maintaining strong coverage.

Using a large-scale proprietary dataset of over \numprint{900000} utterance–product pairs, we observed three main findings. First, while utterance features alone are strong predictors, augmenting them with structured product attributes yields consistent gains. Lightweight summarization methods such as mean pooling and Maximal Marginal Relevance (MMR) improve accuracy without substantial overhead. Second, a precision-oriented auxiliary loss further reduces incorrect visual triggers, which is essential for maintaining a seamless user experience. Third, in backbone comparisons, XLNet achieved the highest predictive performance, whereas distilBERT offered a favorable trade-off between accuracy and efficiency.

These results underline the value of multimodal fusion and task-aligned optimization for cross-device virtual assistants. They also highlight the promise of lightweight generative models, adapted for image-seeking intent prediction, as a practical path toward scalable and responsive production systems. Compact generative models, when paired with effective summarization and loss design, can approach the accuracy of larger architectures while remaining suitable for latency- and resource-constrained environments.

\header{Limitations}  
Our approach is trained using a proxy signal—image carousel interactions—that may not capture all cases of visual intent, leading to potential label noise. The method also depends on structured, high-quality product metadata, which may be unavailable or noisy in other domains. Moreover, while we evaluate multiple transformer backbones, we do not explore recent vision–language or multimodal pretraining approaches that could further improve performance. Finally, our evaluation is offline; live deployment studies are required to assess real-world effectiveness, user satisfaction, and unintended consequences.

\header{Future Work} Future directions include incorporating the extension of the framework to account for sparse multimodal learned retrieval \cite{nguyen2024multimodal, nguyen2024multimodal2}, improving the fusion mechanisms, and collecting explicit human annotations to address the brittleness of the existing retrieval pipeline \cite{hendriksen2025benchmark}. Live A/B testing could provide deeper insights into user experience impacts and system-level trade-offs.

\bibliographystyle{ACM-Reference-Format}
\balance
\bibliography{bibliography}

\end{document}